\documentclass[12pt]{article}
\usepackage{amssymb}
\usepackage{footmisc}
\usepackage{subfigure}
\usepackage{enumitem}
\usepackage[utf8]{inputenc}
\usepackage[margin=1.25in]{geometry}
\usepackage{graphicx}
\usepackage{amsmath}% http://ctan.org/pkg/amsmath
\makeatletter
\newcommand{\xleftrightarrow}[2][]{\ext@arrow 3359\leftrightarrowfill@{#1}{#2}}
\newcommand{\xdashrightarrow}[2][]{\ext@arrow 0359\rightarrowfill@@{#1}{#2}}
\newcommand{\xdashleftarrow}[2][]{\ext@arrow 3095\leftarrowfill@@{#1}{#2}}
\newcommand{\xdashleftrightarrow}[2][]{\ext@arrow 3359\leftrightarrowfill@@{#1}{#2}}
\def\rightarrowfill@@{\arrowfill@@\relax\relbar\rightarrow}
\def\leftarrowfill@@{\arrowfill@@\leftarrow\relbar\relax}
\def\leftrightarrowfill@@{\arrowfill@@\leftarrow\relbar\rightarrow}
\def\arrowfill@@#1#2#3#4{%
  $\m@th\thickmuskip0mu\medmuskip\thickmuskip\thinmuskip\thickmuskip
   \relax#4#1
   \xleaders\hbox{$#4#2$}\hfill
   #3$%
}
\makeatother

\usepackage[notes,backend=biber,autolang=other,bibencoding=latin1,related=true,idemtracker=constrict]{biblatex-chicago}

\usepackage[pdftex,colorlinks,allcolors=black]{hyperref}
\bibliography{key-risks.bib}

\newcommand{\para}[1]{\vspace{2mm}\noindent\textbf{#1.}}
\newcommand{\parait}[1]{\vspace{2mm}\noindent\textit{#1.}}

\title{Bugs in our Pockets:\\The Risks of Client-Side Scanning}
\author{
Hal Abelson \and
Ross Anderson \and
Steven M. Bellovin \and
Josh Benaloh \and
Matt Blaze \and
Jon Callas \and
Whitfield Diffie \and
Susan Landau \and
Peter G. Neumann \and
Ronald L. Rivest \and
Jeffrey I. Schiller \and
Bruce Schneier \and
Vanessa Teague  \and
Carmela Troncoso
}
\date{October 15, 2021}

\begin{document}
\maketitle

\section*{Executive Summary}

% Setting the scene, crypto is seen as a barrier by enforcement
% Enter CSS: what it is and why they claim is good
Our increasing reliance on digital technology for personal, economic, and government affairs has made it essential to secure the communications and devices of private citizens, businesses, and governments. This has led to pervasive use of cryptography across society. Despite its evident advantages, law enforcement and national security agencies have argued that the spread of cryptography has hindered access to evidence and intelligence.  
Some in industry and government now advocate a new technology to access targeted data: \emph{client-side scanning} (CSS). Instead of weakening encryption or providing law enforcement with backdoor keys to decrypt communications, CSS would enable on-device analysis of data in the clear. If targeted information were detected, its existence and, potentially, its source, would be revealed to the agencies; otherwise, little or no information would leave the client device. Its proponents claim that CSS is a solution to the encryption versus public safety debate: it offers privacy---in the sense of unimpeded end-to-end encryption---and the ability to successfully investigate serious crime. 

% Make clear that we doubt both side of the equation. Bring from the beginning that efficacy is a problem
In this report, we argue that CSS neither guarantees efficacious crime prevention nor prevents surveillance. Indeed, the effect is the opposite. CSS by its nature creates serious security and privacy risks for all society while the assistance it can provide for law enforcement is at best problematic. There are multiple ways in which client-side scanning can fail, can be evaded, and can be abused. 

% CSS is effectively a surveillance infrastructure
Its proponents want CSS to be installed on all devices, rather than installed covertly on the devices of suspects, or by court order on those of ex-offenders. But universal deployment threatens the security of law-abiding citizens as well as lawbreakers. Technically, CSS allows end-to-end encryption, but this is moot if the message has already been scanned for targeted content. In reality, CSS is bulk intercept, albeit automated and distributed. As CSS gives government agencies access to private content, it must be treated like wiretapping. In jurisdictions where bulk intercept is prohibited, bulk CSS must be prohibited as well.

% Why CSS does not guarantee privacy
Although CSS is represented as protecting the security of communications, the technology can be  repurposed as a general mass-surveillance tool. The fact that CSS is at least partly done on the client device is not, as its proponents claim,  a security feature. Rather, it is a source of weakness. As most user devices have vulnerabilities, the surveillance and control capabilities provided by CSS can potentially be abused by many adversaries, from hostile state actors through criminals to users' intimate partners. Moreover, the opacity of mobile operating systems makes it difficult to verify that CSS policies target only material whose illegality is uncontested.

% CSS worsens the privacy situation
The introduction of CSS would be much more privacy invasive than previous proposals to weaken  encryption. Rather than reading the content of encrypted communications, CSS gives law enforcement the ability to remotely search not just communications, but information stored on user devices.

%this  And ... CSS is not efficacious 
Introducing this powerful scanning technology on all user devices without fully understanding its vulnerabilities and thinking through the technical and policy consequences would be an extremely dangerous societal experiment. Given recent experience in multiple countries of hostile-state interference in elections and referenda, it should be a national-security priority to resist attempts to spy on and influence law-abiding citizens. CSS makes law-abiding citizens more vulnerable with their personal devices searchable on an industrial scale. Plainly put, it is a dangerous technology. Even if deployed initially to scan for child sex-abuse material, content that is clearly illegal, there would be enormous pressure to expand its scope. We would then be hard-pressed to find any way to resist its expansion or to control abuse of the system.

% Ultimate consequences of opening the door to in-device surveillance
The ability of citizens to freely use digital devices, to create and store content, and to communicate with others depends strongly on our ability to feel safe in doing so. The introduction of scanning on our personal devices---devices that keep information from to-do notes to texts and photos from loved ones---tears at the heart of privacy of individual citizens. Such bulk surveillance can result in a significant chilling effect on freedom of speech and, indeed, on democracy itself.

\newpage
\section{Introduction}
% Setting the scene and the crypto debate
Since strong encryption became available to the public nearly half a century ago, intelligence and law enforcement agencies around the world have sought to limit its use.  The first attempts were to restrict access to encryption devices.  When the invention of the PC made encryption widely available in software, attempts were made to require backdoors that would provide governmental access to decryption. These proposals were effectively abandoned in most democratic countries by the end of the twentieth century. Since then, several national intelligence and law enforcement agencies have instead worked to enlist vendors as surrogates to provide access to encrypted traffic, whether using technical vulnerabilities or covert court-ordered access.

% Moving from removing crypto to client-side scanning
Instead of providing decryption capabilities, many current policy efforts involve trying to circumvent encryption entirely by scanning materials before they are encrypted or after they are decrypted. The leading proposal, \textit{client-side scanning} (CSS), is a phase change in the debate that requires a different analysis from those in prior debates about government access to encrypted data.\footnote{See generally~\cite{KUD} and~\cite{1997Report}, both of which had substantial author overlap with this report.} 

% CSS seems like a great opportunity. Mentioning Apple's positives
At a casual glance, CSS systems may seem an opportunity to provide a compromise approach to surveillance. Data can be encrypted end-to-end in transit and at rest (e.g., in encrypted backup systems), rather than being available in cleartext on services on which governments can serve warrants. Involving the user's device in the CSS process may allow for some sort of transparency; perhaps some cryptography can help verify properties of the scan prior to its execution, or limit the purpose or pervasiveness of scanning. CSS may also allow for some rudimentary user control, as users may be able to decide what content can be scanned, or remove the scanning altogether. Some of these properties are, for instance, provided by Apple's August 2021 proposal, which was advertised as limited in scope; looking only for images of abuse that have been certified as illegal under international treaty; as having elaborate cryptographic mechanisms to prevent leakage of non-targeted material; and allowing the user to avoid scanning by stopping their Camera Roll being backed up to iCloud. 

% No privacy
However, when analyzing CSS systems---including Apple's proposal---from a security perspective, it becomes apparent that the promise of a technologically limited surveillance system is in many ways illusory. While communications can be encrypted, users' data is still searched and scrutinized by law enforcement in ways that cannot be predicted or audited by the users. This leads to some obvious questions: How is the list of targeted materials obtained?  What prevents other materials from being added to the list, such as materials that are lawful but that displease the government of the day? How and to whom is the discovery of targeted materials reported? What safeguards protect user privacy and keep third parties from using these channels to exfiltrate data?

% And expanded surveillance
Existing device scanning products such as antivirus software and ad blockers act to protect the user. By contrast, CSS acts against the user. Its surveillance capabilities are not limited to data in transit; by scanning stored data, it brings surveillance to a new level. Only policy decisions prevent the scanning expanding from illegal abuse images to other material of interest to governments; and only the lack of a software update prevents the scanning expanding from static images to content stored in other formats, such as voice, text, or video.

% and no efficacy
In addition to the fundamental questions of whose interests are being served and whether privacy and purpose limitation can be enforced, there are also technical questions around whether CSS can actually be a safe and effective tool to detect crime. Are the algorithms to detect targeted content robust against adversarial modifications? Can adversaries influence the algorithms to avoid detection? Can adversaries use the detection capabilities to their advantage (e.g., to target opponents)?

In this report, we explore possible answers to these questions. In the end, we find no design space for solutions that provide substantial benefits to law enforcement without unduly risking the privacy and security of law-abiding citizens.

This report builds on the work of many others. We build on recent work by the US National Academies of Science, Engineering, and Medicine, which provides a framework for evaluating policy or technical approaches for access to unencrypted content~\autocite{NASEM2018}, and on the 2019 Carnegie Endowment for International Peace study on encryption policy, which presents a set of principles with which to guide solutions \autocite{CEIP2019}. We also build on Paul Rosenzweig's early analysis of the policy issues raised by CSS, along with some of the technical issues~\autocite{Rosenzweig2020}. Since Apple announced its scanning proposal in August 2021, several researchers and organizations have provided rapid analyses of that proposal, and the technology and policy issues it raises. In particular, we acknowledge Eric Rescorla of Mozilla~\autocite{Rescorla2021}, 
Kurt Opsahl of the Electronic Frontier Foundation~\autocite{Opsahl2021}, Steven Murdoch~\autocite{Murdoch2021}, Paul Rosenzweig~\autocite{Rosenzweig2021}, and Daniel Kahn Gillmor of the ACLU.~\autocite{Gillmor2021} Here our aim is a more thorough technical analysis, and to cover CSS more generally. In addition, our analysis sheds some light on the design decisions that Apple took. Apple did its best, using some of the top talent in security and cryptography, and yet did not achieve a design for a secure, trustworthy, and efficacious system.

\para{Terminology} In what follows, we refer to text, audio, images, and videos as ``content,'' and to content that is to be blocked by a CSS system as ``targeted content.'' This generalization is necessary. While the Five Eyes governments and Apple have been talking about child sex-abuse material (CSAM)---specifically images---in their push for CSS~\autocite{5eyes2020}, the European Union has included terrorism and organized crime along with sex abuse.~\autocite{Coreper2020} In the EU's view, targeted content extends from still images through videos to text, as text can be used for both sexual solicitation and terrorist recruitment. We cannot talk merely of ``illegal'' content, because proposed UK laws would require the blocking online of speech that is legal but that some actors find upsetting.\autocite{onlinesafety2021} 

Once capabilities are built, reasons will be found to make use of them. Once there are mechanisms to perform on-device censorship at scale, court orders may require blocking of nonconsensual intimate imagery, also known as revenge porn. Then copyright owners may bring suit to block allegedly infringing material. So we need to understand the technological possibilities of existing and likely near-future content-scanning technologies.

One issue of definition is what counts as a client and what is a server. Obviously, one's own devices, such as desktop and laptop computers, smartphones, and tablets, are clients. Also obviously, social media and content-sharing websites are servers. It is beyond the scope of this report to completely explore the gray area in between the two. However, a useful guideline for the grey area is that of a private space as opposed to a public space. A personal file server in one's own home, which exists to augment a user's private storage, resembles the internal storage of a laptop more than it resembles social media. An end-to-end encrypted messenger application is also a private space. Yet there are many systems in the middle: cloud backup, cloud drives, generic encrypted storage hosted in the cloud, and so on. Our analysis centers around whether the system is intended to be a private space, in which case we consider it to be a client regardless of the underlying technology. On the other hand, we would consider a blog to be a server, even if it is personally hosted, since it is generically available on the Internet.

\section{Content Scanning Technologies}
\label{sec:content_scanning}

Many online service providers that allow users to send arbitrary content to other users already perform periodic scanning to detect objectionable material and, in some cases, report it to authorities. Targeted content might include spam, hate speech, animal cruelty, and, for some providers, nudity. Local laws may mandate reporting or removal. For example, France and Germany have for years required the takedown of Nazi material, and the EU has mandated that this be extended to terrorist material generally in all member states. In the US, providers are required to report content flagged as CSAM to a clearinghouse operated by the National Center for Missing and Exploited Children (NCMEC), while in the UK a similar function is provided by the Internet Watch Foundation (IWF). 

Historically, content-scanning mechanisms have been implemented on provider-operated servers. Since the mid-2000s, scanning has helped drive research in machine-learning technologies, which were first adopted in spam filters from 2003. However, scanning is expensive, particularly for complex content such as video. Large machine-learning models that run on racks of servers are typically complemented by thousands of human moderators who inspect and classify suspect content. These people not only resolve difficult edge cases but also help to train the machine-learning models and enable them to adapt to new types of abuse.

One incentive for firms to adopt end-to-end encryption may be the costs of content moderation. Facebook alone has 15,000 human moderators, and critics have suggested that their number should double.\autocite{Barrett2020} The burden of this process is much reduced by end-to-end encryption as the messaging servers no longer have access to content. Some moderation is still done based on user complaints and the analysis of metadata. However, some governments have responded with pressure to re-implement scanning on user devices. 

In this section, we summarize the current technical means to implement scanning, and explore the difference between deploying such methods on the server or on the client.

\subsection{Content Scanning Methods}

Currently, two different technologies are used for image scanning: perceptual hashing and machine learning.

\para{Perceptual hashing}
Hashes are specialized algorithms capable of digesting a large input file and producing a short unique ``fingerprint'' or hash. Many scanning systems make use of \textit{perceptual hash functions}, which have several features that make them ideal for identifying pictures. Most importantly, they are resilient to small changes in the image content, such as re-encoding or changing the size of an image. Some functions are even resilient to image cropping and rotation. 

Perceptual hashes can be computed on user content and then compared to a database of targeted media fingerprints, in order to recognize files that are identical or very similar to known images. The advantage of this approach is twofold: (1) comparing short fingerprints is more efficient than comparing entire images, and (2) by storing a list of targeted fingerprints, providers do not need to store and possess the images. 

In the case of child sex-abuse images, it is an offense in many jurisdictions to possess such material; providers who come across it are required to report it immediately and destroy it. National abuse organizations such as NCMEC in the USA and the IWF in the UK receive these reports and have legal cover to retain and curate such material. Service providers therefore use a list of hashes of images assembled by these organizations. Examples of perceptual hash functions include phash, Microsoft's PhotoDNA~\autocite{photodna}, Facebook's PDQ hash function~\autocite{Newton2019Facebook-open-s}, and Apple's NeuralHash function.~\autocite{apple-childsafety}

\para{Machine Learning} The alternative approach to image classification uses machine-learning techniques to identify targeted content. This is currently the best way to filter video, and usually the best way to filter text. The provider first trains a machine-learning model with image sets containing both innocuous and target content. This model is then used to scan pictures uploaded by users. Unlike perceptual hashing, which detects only photos that are similar to known target photos, machine-learning models can detect completely new images of the type on which they were trained. One well-known example is the face detector used in iPhones to detect faces on which to focus the camera. 

\bigskip
While these two scanning technologies operate differently, they share some common properties. Both require access to unencrypted content for matching. Both can detect files that the system has not seen before, though perceptual hashing is limited to detecting files that differ only slightly from images it has seen before. Both methods have a non-zero false positive rate. Both methods also rely on a proprietary tool developed from a corpus of targeted content, which may be controlled by a third party. Some scanning techniques also use proprietary algorithms (for example, Microsoft's PhotoDNA is available only under a nondisclosure agreement). Finally, regardless of the underlying technology, either method can be treated as a black box that inputs an unencrypted image and outputs a determination of whether it is likely to contain targeted material. 

These commonalities result in scanning based on both methods having very similar security properties. Both methods can be evaded by knowledgeable adversaries (see \autoref{sec:effectiveness_attacks}); and both methods can be subverted in similar ways (see \autoref{sec:functionality}).

\subsection{Content Scanning Operation Flows}\label{sec:flows}

As we explain in the previous section, scanning can be used to discover whether the user has content on a known-bad list (e.g., using perceptual hashing to find whether the user has a known CSAM image) or used to find content of a target class (e.g., by using machine learning). We now describe the flow of actions to perform scanning at the server and at the client, respectively.

\para{Scanning at the Server} Currently, most tech companies' scanning processes run on their own servers. There are two reasons for this. First, the companies agree to host or transmit only certain types of material, and customers must consent to this when they register for the service. Firms then search customer data to prevent the sharing of material that is illegal or against their terms of service (e.g., in Facebook's case, nudity). Second, it is convenient: customer data is easily accessible on their servers, which have the computational capacity for the task. As technology has evolved, people do not always send images and other content to each other directly, but rather send links to material stored in the cloud. This enables tech companies to search shared images~\autocite{Negreiro2020} and means that the tech companies are not typically examining material that is privy to a single account or individual. 

\begin{figure}
    \centering
    \includegraphics[width=\textwidth]{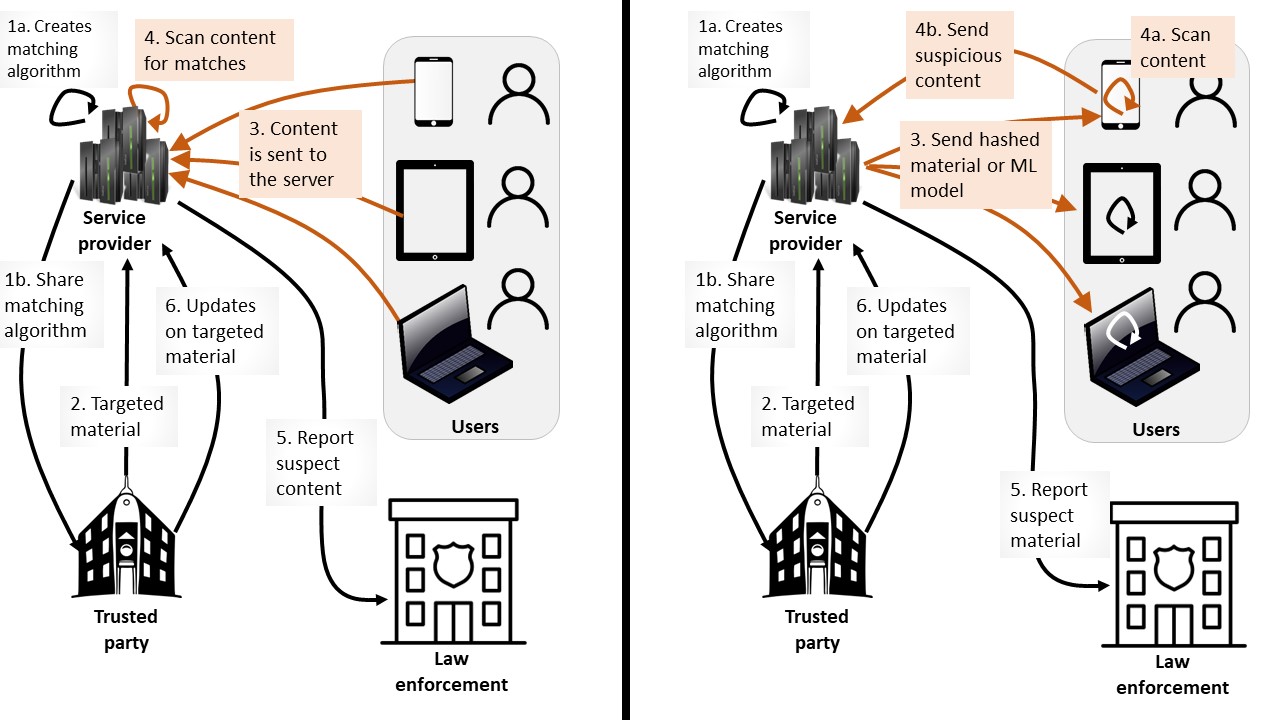}
    \caption{Scanning operation flows. \textbf{Left}: Server-side scanning.
    \textbf{Right}: Client-side scanning (the main changes are in orange)}.
    \label{fig:css_arch}
\end{figure}

We will now consider the operational flow, starting with the case of server-side scanning using perceptual hashing, illustrated in \autoref{fig:css_arch} (Left):

\begin{enumerate}
    \item The service provider selects or creates a perceptual hashing algorithm, and provides it to the curator of targeted material (e.g., NCMEC or IWF).
    \item The curator returns a list of the hashes of targeted material.
    \item Users send content to or through the server.
    \item The server scans all uploaded content, looking for targeted content. Suspect material can be sent for adjudication to human reviewers.~\autocite{EU_Tech_Sol_2020} 
    \item Depending on the type of material and local law, verified positive matches may also be reported to law enforcement agencies for further action. This will vary by jurisdiction and application. 
    \item The hash list is periodically updated by the curator.\footnote{The frequency of updates will be dictated by the ability of producers of targeted content to react to scanning, and the ability of the curator to react to the producers. In a contested environment, as with email spam, this may mean daily updates.}  
\end{enumerate}
If machine learning were used instead of perceptual hashing, in step 1 the service provider would select a model architecture and training algorithm; in step 2, the curator would train the model with their targeted material; and in step 4 the scan would apply the model to content sent by users.

Implementing a scanning system within a centralized online server backed up with human reviewers has benefits. First, the provider does not need to publish their algorithms or model, and is not constrained in terms of computation. Second, server-side scanning allows the provider to use techniques such as clustering across large numbers of accounts simultaneously to identify similar content and make detection more reliable. For instance, spam detectors learn to identify spam messages by relying on reports from millions of email users clicking the ``report spam'' button to retrain the spam filter models every day. 

\para{Scanning at the Client} In a generic CSS system, every relevant device---including phones, computers, tablets, and perhaps even watches and speakers---would have software to monitor activity and alert authority if the user acquired targeted material or sent it to another user. This could circumvent encryption completely by monitoring all content prior to the user encrypting and sending it, or after receiving it, or when backing it up to the cloud. It could even monitor notes that the user wrote for their own use with no intention of ever backing them up or sending them to anybody else. It would thus replicate the behavior of a law-enforcement wiretap. Although various intelligence and law-enforcement agencies have been lobbying for CSS since 2018, Apple has been the first to propose a concrete design.~\autocite{apple-childsafety} We analyze the security implications of Apple's design in \autoref{sec:apple}.

The operational flow of client-side scanning, illustrated in ~\autoref{fig:css_arch} (Right), is very similar to that of server-side scanning:

\begin{enumerate}
    \item The CSS service provider selects or creates a matching algorithm, and provides it to the curator of targeted material (e.g., NCMEC or IWF for perceptual hashes for CSAM).
    \item The curator returns a list of the hashes of targeted material.
    \item The hash list is incorporated into production code and installed in users' devices according to the normal update cycles, such as Windows Update or Apple's System Update.
    \item The CSS runs on the user device looking for targeted content. Because of the scale and complexity of scanning, suspect material is likely to have a second, automated, scan on the server and then, potentially, a human review.~\autocite{EU_Tech_Sol_2020} 
    \item Depending on the type of material and local law, verified positive matches may also be reported to law enforcement agencies for further action. This will vary by jurisdiction and application.
    \item The hash list is periodically updated by the curator and pushed to client devices through the update cycle. 
\end{enumerate}

In some CSS variants, the system may notify other parties instead of law enforcement. For instance, after targeted content is found in step 4, instead of sending it to the server, or notifying law enforcement (step 5), the CSS system may launch a local notification to prevent the user from performing an action or to ask them to reconsider. Alternatively, the CSS system may notify others such as the parent or guardian of a child, if the device is owned by a minor.~\autocite{apple-childsafety} Our analysis in the following sections holds independently of who the CSS system notifies upon finding a match.

CSS is designed to be similar to server-side scanning---out of the user's control, and searching everything without either a warrant or individualized suspicion. Yet it lacks some of the advantages of server-side scanning. At least some part of the scanning algorithm has to be run on the client, with the consequent danger of being made public together with the targeting data, such as a list of hashes or an ML model. If it used an ML model rather than perceptual hashing, there would be an elevated risk that an adversary could perform a model-extraction attack, or even extract some of the training data; we will discuss this in more detail later in Section~\ref{sec:privacyrisks}.

In CSS, the provider is also constrained in terms of computation and data. For instance, consider the EU's demand to scan text messages to detect serious crimes, including grooming and terrorist recruitment. At present, Europol uses keyword scanning, with a list of some 5,000 words in multiple languages---including slang---for drugs and guns. If such method were deployed on user devices at scale, it would presumably report anyone that had more than a certain threshold of these words in their messages. This would lead to many false positives among law-abiding hunters, gun collectors, writers and the like. CSS cannot rely on the large-scale clustering techniques used by modern spam filters; and determining topic and intent in large corpora of text are hard problems. 

\section{Security and Policy Principles for Content Scanning}
\label{sec:principles}

Moving scanning from the server to the client pushes it across the boundary between what is shared (the cloud) and what is private (the user device). By creating the capability to scan files that would never otherwise leave a user device, CSS thus erases any boundary between a user's private sphere and their shared (semi-)public sphere.~\autocite{Rescorla2021} It makes what was formerly private on a user's device potentially available to law enforcement and intelligence agencies, even in the absence of a warrant. Because this privacy violation is performed at the scale of entire populations, it is a bulk surveillance technology. 

Different jurisdictions have different tests for intrusions on fundamental rights. In Europe, the test is whether the intrusion is not just in accordance with law but also ``necessary and proportionate.'' To understand whether CSS can be justified as such, we need to look at the nature of its actual and reasonably foreseeable intrusions in detail. Judicious and responsible choices about surveillance technologies should be founded on the technical principles of security engineering, and on the policy principles that reflect a society's values. In examining the technical and societal risks of CSS, there are several distinct concerns: how to ensure that information within the system is properly protected (a security concern); how to ensure that the system of central servers, human reviewers, user devices, users, and potentially targeted content works appropriately (a socio-technical concern); and how to ensure that technologies with potential to become bulk-surveillance infrastructure can be deployed safely (a policy concern).

In this section, we identify the threats from which CSS systems should protect users, and the security and policy design principles that could enable them to do so.

\subsection{Threats to CSS Systems}
\label{sec:threats-to-css}

\begin{figure}
    \centering
    \includegraphics[width=0.85\textwidth]{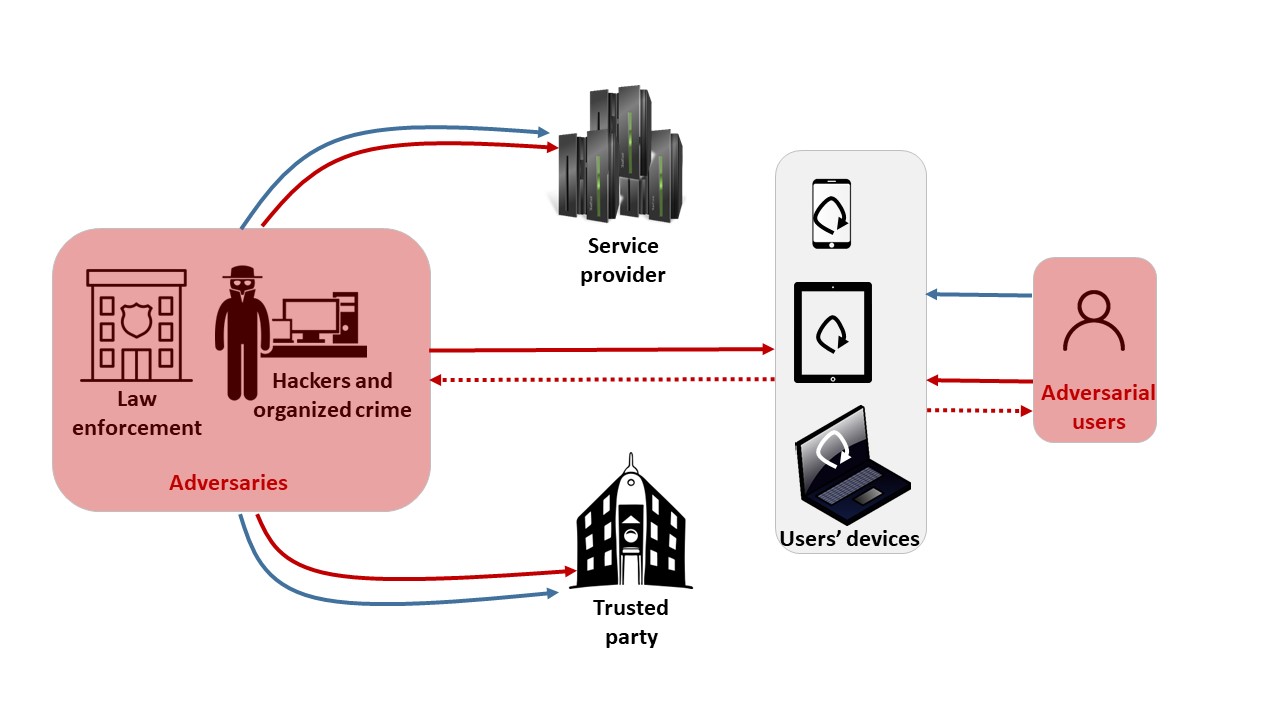}
    \caption{From server-side to client side: New compromise paths and advantage points for adversaries (\textcolor{blue}{$\longrightarrow$}: compromise paths in server-side scanning;
    \textcolor{red}{$\longrightarrow$}: compromise paths in CSS;
    \textcolor{red}{$\xdashrightarrow{\textcolor{white}{lllll}}$}: knowledge gained by adversary in CSS) }.
    \label{fig:adversaries}
\end{figure}

Security must be defined with respect to the threats that the security engineer anticipates. A first observation is that moving content scanning capabilities from the server to the client opens new vantage points for the adversary. We illustrate this extension of the attack surface in \autoref{fig:adversaries}. Attacks that already existed on server-side scanning can now be executed by more actors, and on less-secure infrastructure (users' devices rather than corporate servers). Moreover, new on-device attacks become possible. 

These new vantage points can be exploited not only by device owners, but also by third parties such as foreign government agencies and criminals. We classify the threats to CSS systems in three groups: abuse by authorized parties such as governments; abuse by unauthorized parties such as dishonest government officials or service provider staff, or equivalently by outsiders who can hack into their systems; and abuse by local adversaries such as abusive partners or controlling family members.

% \begin{enumerate}
%     \item The first is authorized access and its abuse, such as governments ordering service providers to modify for more types of targeted content, or directing the mechanisms at political opponents;
%     \item The second is unauthorized access, such as by tech company staff or hacking by foreign states;
%     \item The third is that CSS may facilitate new types of harm, such as if parents in child-custody disputes can use it to make false allegations of abuse, or if children can make false allegations against each other, or if the threat of such allegations can be used to discriminate against LGBTQ kids.
% \end{enumerate}

\para{Abuse by Authorized Parties} Many critics of CSS have pointed out that governments have the power to order the service provider to search for material beyond the initial, publicly acknowledged target. The European Union has been pushing for CSS since July 2020, with concerns  not just about child abuse, but also terrorism.~\autocite{EU_Tech_Sol_2020} Australian police have raided journalists for publishing images of war crimes; and the UK seeks to ban a range of online content, including some forms of speech that are perfectly legal face-to-face.\autocite{onlinesafety2021} All three already have laws enabling them to mandate tech companies to repurpose or retarget existing scanning techniques. Once we move beyond the OECD countries, the legal constraints against government action become weaker and the range of targeted content becomes larger, typically including LGBTQ+ content, political activists, and domestic rivals of authoritarian regimes\autocite{Higgins2016How-Moscow-Uses}. In such places, CSS will provide a means of repression and political manipulation. 

Augmenting the scope of CSS may not only refer to topic, but also to algorithmic changes. For instance, given that NCMEC's mission includes missing children, it would not be surprising if there were pressure for CSS to be augmented with a face-recognition capability.

\para{Abuse by Unauthorized Parties}
Additional threats stem from unauthorized access, such as second-party (corrupt police officers or tech company staff) and third-party (foreign state or criminal) hacking. CSS makes surveillance systems more complex than its server-side predecessors; by expanding the attack surface, it creates new points of technical failure and more powerful insiders who might be subverted, coerced, or hacked. 

This threat overlaps with the previous: for example, a corrupt police officer may be working not just for organized crime but for a foreign state. Another overlap is in supply-chain attacks. One example: for decades, the US and German intelligence services covertly owned the Swiss company Crypto AG---the main provider of cryptographic equipment to non-aligned countries. This operation gave NATO countries privileged access to much of the world's diplomatic and military traffic. Similarly, there have been concerns that an antivirus software vendor might have a covert relationship with a national intelligence service. If CSS were to become pervasive, there would be an enormous incentive for nation-states to subvert the organizations that curated the target list, especially if this list were secret. Equivalently, they might try to suborn individuals within these organisations, or hack their computers. In effect, the whole supply chain for samples of targeted content comes within the trusted computing base of everyone whose device uses the CSS system.

We also note that when scanning is performed mostly at the client side,
attacks would more likely be conducted on citizens' devices rather than on a central organizational asset such as a messaging or filtering server. Therefore, service providers may be both less able and less motivated to defend against attacks.

\para{Local Adversaries}
Another class of threats comes from local adversaries, such as the user's partner, ex-partner, other family member, or personal rival. Consider, for example, a woman planning to escape a violent or controlling partner who abuses both her and their children. The abuser will often install stalkerware on family phones, and he may use ``smart home'' infrastructure such as video doorbells as a way to control them.~\autocite{Levy2020Security--Threa} The same happens with child abuse. The great majority of threats to children, whether sexual, physical, or emotional, come not from strangers but from members of their social circle, including family members, friends, and members of their class at school. 

Technical security mechanisms, including CSS systems, are not well designed for such cases; much of the standard security advice and practice is ineffective or even counterproductive. Advising victims to change their passwords is of little help when the abuser knows the answers to their security questions. In the related topic of child safety, system designers must carefully consider many issues from school bullying to queer kids who need privacy.  

Child protection and misogynistic violence are too complex to be reduced to a simple arithmetic of ``thirty strikes and you're out.'' Not all alarms indicate the same level of abuse, and not all require the involvement of law enforcement. Only 1\% to 2\% of child-welfare cases indicate a child-protection case, where a child is an imminent risk of serious harm.\autocite{ABCDKM2006} Automated scanning will likely result in so many reports that service providers or law enforcement will need to triage manually. Without human screening, if the bar is set high enough to keep the volumes manageable, then most of the serious alarms will be missed.\footnote{Last year, Facebook reported over 20 million cases of suspected CSAM in the USA alone. Unlike with wiretapping, proper statistics are not available on outcomes. If the FBI or NCMEC wishes to extend scanning to clients, then it is reasonable to ask them to publish figures on how many cases went to what stage of investigation, including prosecutions and convictions.} And for every urgent child protection case, there will be dozens to hundreds of cases that require more sensitive handling. Above all, systems must respect children's rights, and this especially applies to systems promoted for child protection.

\subsection{Core Security Engineering Principles}
\label{sec:secengprin}

Good security engineering starts with a realistic threat model that explains who is likely to attack a system and why. This serves as the basis for a security policy that sets out what actions are to be prevented or detected. The security policy is then implemented using a variety of protection mechanisms, which may include access controls, cryptography, and alarms. The final piece is assurance, a process whereby the stakeholders verify the design, validate the implementation, and monitor the operation. The process is not purely technical, as the security engineer must pay attention to who has an incentive to protect the system and who suffers when it fails. 

\para{Security Engineering Best Practices} The security engineer must understand how such threat models, security policies, and protection mechanisms have failed in the past and then take steps to manage the residual risk.\autocite{SEv3}

To illustrate this point, let us take three relevant examples of security policies that, like CSS, (i) have as the main threat authorized insiders that could abuse the system; and (ii) use \textit{mandatory access controls}---software mechanisms that enforce certain properties regardless of user actions.

First, security policies originating in the US Department of Defense (DoD) aim to ensure that information could flow upwards from unclassified to Secret (or Top Secret), but not downwards.\footnote{These are known as \textit{multilevel security} policies} Both intelligence and law enforcement agencies want surveillance to remain secret from its targets; nobody should know whether they have been wiretapped until they are arrested and confronted with the evidence.\footnote{In certain cases, e.g., in wiretaps conducted under the US Foreign Intelligence Surveillance Act, the suspect may never learn that they have been surveilled.} Workstations used by intelligence officers ensure that Secret documents are not mailed outside of the organization by mistake. Above all, nobody should be technically able to make a bulk disclosure of highly classified information. 

Second, commercial security policies seek to guarantee that insiders cannot commit fraud: for example, by ensuring that transactions of consequence are authorized by more than one officer; and by ensuring that money cannot be created or destroyed, merely moved from one account to another. Here, too, no individual should have the ability to do so much damage that the firm is destroyed.

Third, smartphone-oriented security policies aim to ensure that hostile apps cannot interfere with other apps or steal information from apps or the phone platform itself. Mobile operating systems such as iOS or Android therefore prevent applications accessing each others' storage or running memory. The protections in iOS are more thorough and make it much harder to deploy stalkerware than is the case for Android.

These three policies focus on preventing harms stemming from insider threats (intelligence or law enforcement agents, bank employees, or installed apps). When they have failed in the past, it was not because the software failed, but because a trusted insider operated contrary to the approved policy. Edward Snowden was able to leak thousands of top-secret NSA documents because he was one of the people entrusted with maintaining the agency's technical security. The rogue trader Nick Leeson was able to destroy Barings Bank because he appeared to make most of the bank's profits, so executives trusted him. And the French and Dutch police were able to take over Encrochat, a chat app used by drug dealers, when they subverted the app's trusted update server and got it to push a doctored version of the app to make the traffic easy to intercept.

A main reason for such failures is that the systems that these policies aim to protect are complex. As such, their \textit{trusted computing base}---the set of hardware, software, and people who can break the security policy---is quite large. This includes not just the mandatory access controls or cryptographic mechanisms used to enforce the security policy, but also the powerful insiders who by accident or design can compromise it. 

The threat of insiders applies to CSS systems. As the  examples above have shown, preventing CSS systems spectacularly failing requires sharply limiting the potential damage that insiders can do. This means not relying on just one actor or component for security-critical actions, and  minimizing the trusted computing base on which the  CSS rests, as well as keeping humans out of it to the extent possible. A main goal is that disloyalty or incompetence of some actor or small group of actors should  not have catastrophic consequences. 
% and that the inevitable failures will be detected in time for appropriate action to be taken.

\bigskip
\para{Design Principles} 
In addition to the security engineering practices that have evolved since the 1960s in government, commerce, and the computer industry, there has been substantial academic research in computer security since the 1970s. In a seminal 1975 paper, Jerry Saltzer and Mike Schroeder proposed eight design principles for protection mechanisms.~\autocite{Saltzer1975The-protection-} These principles are particularly relevant to content-scanning systems:
\smallskip

\noindent\textit{Economy of mechanism}, which states that the design should be as simple and small as possible, so it is easy to validate and test. (This is their equivalent of the DoD principle that the trusted computing base be minimized.)

\noindent\textit{Fail-safe default}, which states that access decisions should be based on permission rather than exclusion.

\noindent\textit{Separation-of-privilege}, which states that where possible, security should rely on more than one entity. (This is their equivalent of the commercial principle of dual control of critical transactions.)

\noindent\textit{Least-privilege}, which states that every program and every user of the system should operate using the least set of privileges necessary to complete the job.

\noindent\textit{Least-common-mechanism}, which states that users should only share system mechanisms when necessary, because shared mechanisms can provide unintended communication paths or means of interference.

\noindent\textit{Open design}, which states that design should not be secret. (This goes back to Auguste Kerckhoffs' 1883 principle in cryptography.~\autocite{kerckhoffs1} As Claude Shannon later put it, it is prudent to assume that ``the enemy knows the system"; thus security lies entirely in the management of cryptographic keys.~\autocite{shannon})

\noindent\textit{Psychological-acceptability}, which states that the policy interface should reflect the user's mental model of the system. Users won't use protection correctly if the mechanics don't make sense to them.

\smallskip
Many of these principles were inspired by early defense research and found their way into secure designs for defense, commercial, and consumer computer systems. Following these principles helps minimize the probability of a security breach. We have already discussed the need to minimize the set of humans and technological components that must be trusted in order to avoid catastrophic consequences, including not relying on a single actor for particularly critical controls. 

In addition, all security-relevant design details should be open and auditable. This includes algorithms used for decision making; when possible, these algorithms should be diversified across users so that a single failure does not become a vector for an attack at scale. And whenever detection algorithms fail---whether accidentally or under attack---they should do so in a way that is safe for the users. Finally, the detection mechanisms should be not just clear to the users, but align with users' mental models, so that users can understand how they are protected and how their actions may affect this.

\bigskip
\para{Damage Control through Purpose Limitation} Systems with security or safety requirements are also designed to limit the harm that they can do, including harm to third parties. In the case of surveillance systems, one strategy is to limit them to specific purposes, such as speed cameras that only detect vehicles breaking the speed limit at one location, and the software in scanners and copiers that prevents the copying of banknotes. The other is to limit them to specific targets, as with the law enforcement interfaces for wiretapping mobile phone systems that support only a limited number of wiretaps, which are supposed to be used only following an appropriate judicial order, and whose use can be audited.

CSS must not be an exception to the principle of purpose limitation. Scanning systems should be designed so as to limit their scope in terms of what material can be scanned (topic and format), as well as where (which components of memory are accessible to the scanner), and by whom (who sets targets and receives alerts). 

\subsection{Core Policy Principles}
\label{sec:carnegie}

Nations that enjoy democracy and the rule of law restrict some executive actions, and have long restricted state surveillance powers by prohibiting some forms of surveillance outright\footnote{In the US, for example, bulk surveillance of the content of domestic communications would violate the Fourth Amendment, while in the EU bulk surveillance without warrant or suspicion is against the Charter of Fundamental Rights.} and restricting others through warrant mechanisms. How would such policies look like for client-side scanning systems?

For guidance in this complex environment, we turn to a 2019 study on encryption policy by the Carnegie Endowment for International Peace.~\autocite{CEIP2019} Though the study's focus was on forensic access to mobile phones, the set of principles it proposed to guide government policy and legislation has immediate relevance to our topic.\footnote{This study involved former senior administrative leadership in national security and law enforcement and members of industry, academia, and civil-society organizations; members included Jim Baker, former General Counsel, FBI; Tom Donahue, former Senior Director for Cyber Operations, National Security Council; Avril Haines, former Deputy Director, CIA, former Deputy National Security Advisor, and current Director of National Intelligence; Chris Inglis, former Deputy Director, NSA; Jason Matheny, former Director, IARPA; Lisa Monaco, former Assistant to the President for Homeland Security and Counterterrorism; as well as two members of this group, Susan Landau and Ronald Rivest.}

The relevance of these principles extends beyond the United States. In the European Union, for example, rights such as privacy are entrenched in the Charter of Fundamental Rights. Such rights can be overridden only by law and in ways that are both necessary and proportionate. Infringements of rights must be limited and purpose-specific, which is aligned with the principles laid out in the Carnegie study.

We list the Carnegie principles here, explaining how each one applies to operational requirements on CSS technologies.

\parait{Law Enforcement Utility} ``The proposal can meaningfully and predictably address a legitimate and demonstrated law enforcement problem.''

To fulfill this principle, CSS would need to be designed in such a way that, given a demonstrated law enforcement problem, the scanning technology would have a low rate of false negatives (e.g., traffickers in targeted material who are not identified by scanning) and false positives (e.g., users being wrongly flagged for possessing such materials). To be effective, it would have to detect video as well as static images, as for some years most child sex-abuse material has been in the form of  video~\autocite{Bursztein2019}.

\parait{Equity} ``The proposal offers meaningful safeguards to ensure that it will not exacerbate existing disparities in law enforcement, including on the basis of race, ethnicity, class, religion, or gender.''

To fulfill this principle, CSS would need to operate in such a way that errors in the decision-making process do not disproportionately affect minorities. This includes, among others, errors due to algorithmic bias, which are known to increase racial inequality~\autocite{BuolamwiniG18}, and errors due to a lack of context when making decisions that can create major disadvantages, e.g., for queer kids.~\autocite{Redmiles21}

\parait{Authorization} ``The use of this capability on a phone is  made available [only] subject to duly authorized legal processes (for example, obtaining a warrant).'' 

CSS technology can be designed in two ways: either it scans all communications and/or stored files of certain types, or it scans only in cases where there is court authorization.\footnote{Apple's system, for example, scans all iMessage communications for children who are part of a Family Sharing Plan and all images that are destined for iCloud Photo.} The latter is more consistent with privacy and human-rights law.

To fulfill this principle, CSS would need to be designed in such a way that the scanning is activated on a device only after due process. Therefore, a CSS system set up to scan all devices fails this principle. 

\parait{Specificity} ``The capability to access a given phone is only useful for accessing that phone (for example, there is no master secret key to use) and that there is no practical way to repurpose the capability for mass surveillance, even if some aspects of it are compromised.''

To fulfill this principle, CSS would need to be designed in such a way that the mechanism to scan one device could not be used for scanning all devices. It really matters whether a CSS technology ships as part of a system that everyone uses (whether an app or a platform) or is installed only following a court order. A CSS system installed on all devices is one easily repurposed for mass surveillance; this does not fulfill the specificity principle.

\parait{Focus} ``The capability is designed in a way that it does not appreciably decrease cybersecurity for the public at large, only for users subject to legitimate law enforcement access.''

CSS systems that are universally deployed, even if they are dormant pending a warrant, introduce risks for all users, as they can be exploited by a variety of stakeholders (see \autoref{sec:functionality}). Thus, any universal deployment of CSS would violate the focus principle. 

\parait{Limitation} ``The legal standards that law enforcement must satisfy to obtain authorization to use this capability appropriately limit its scope, for example, with respect to the severity of the crime and the particularity of the search.'' 

To fulfill this principle, CSS must be strictly limited in purpose, or deployed only on the devices of individuals against whom the authorities have obtained proper legal authorization. CSS architecture is such that governments could easily coerce its use for much broader purposes; and it is designed to run on all devices rather than just those of a handful of suspects.

\parait{Auditability} ``When a phone is accessed, the action is auditable to enable proper oversight, and is eventually made transparent to the user (even if in a delayed fashion due to the need for law enforcement secrecy).''

To fulfil this principle, CSS would need to be designed in such a way that users could eventually learn which content was scanned, which content was determined to be targeted and what was ultimately made available to the authorities. Additionally, CSS would need to be auditable; that is, it should be possible to know what content the scanning technology has been targeting. This means that the target images used to create hash lists, or the target training data used to evolve neural-network models, would need to be made available to knowledgeable parties who are able and willing to mount an adversarial challenge if need be. If the target samples are illegal, then a legal safe haven must be created for audit, just as many countries create a safe haven for list curation.

\parait{Transparency, Evaluation, and Oversight} ``The use of the capability will be documented and publicly reported with sufficient rigor to facilitate accountability through ongoing evaluation and oversight by policymakers and the public.''

To fulfill this principle, the complete CSS design (all systems, protocol, and algorithmic aspects) would need to be made public prior to deployment, in such a way that its operation is reproducible and can be evaluated publicly. Apple deserves praise for doing this, and for delaying the launch of its system following the publication of preimage and adversarial attacks on the NeuralHash algorithm. 

A fundamental tenet of the Carnegie principles is that surveillance technologies must have sufficient technical and policy controls that they cannot be repurposed in ways that decrease the safety and security of non-targeted individuals or of society at large. Because CSS technology has the potential to scan data anywhere on a device, such controls are critical. 

\section{CSS Introduces New Privacy and Security Risks}
\label{sec:functionality}
%\section{Client-Side Scanning: Expanding the Risk Surface}\label{sec:css}

In this section, we analyze the extent to which CSS technologies increase the security and privacy risks for users with respect to the status quo in which there is no scanning in the device, and some service providers perform scanning on the material users upload to the server. 
The risk increase stems mainly from the fact that CSS introduces new background software on users' devices. This process does not provide any benefit to the user, but is there to work against the user's interests, and in many cases without their knowledge and consent.

\subsection{Privacy Risks}
\label{sec:privacyrisks}

We first analyze the increase of privacy risks: how much information can an adversary learn about a target by subverting a CSS system? What new privacy harms are brought by CSS systems?

\para{Expansion of scanning to other targets and system components}
Before diving into the ways in which the introduction of CSS may enable adversaries to learn information about users, we first show how dangerously it augments the monitoring capabilities of any party---authorized or not---that can exploit the system. The scanning of clients rather than of communications means that such a party gains capabilities that do not exist at present, and that would not have been created by any previous proposal to weaken or backdoor encrypted communications.

When scanning for targeted material takes place at a central server, it affects only content the user uploads for sharing with others, such as photos and emails. The same communications monitoring capabilities would also have been provided by previous proposals for banning encryption or introducing government backdoors in encryption algorithms or protocols.

The deployment of CSS changes the game completely by giving access to data stored on users' devices. First, it facilitates global surveillance by offering economies of scale. Second, while proposals are typically phrased as being targeted to specific content, such as CSAM, or content shared by users, such as text messages used for grooming or terrorist recruitment, it would be a minimal change to reconfigure the scanner on the device to report any targeted content, regardless of any intent to share it or even back it up to a cloud service. That would enable global searches of personal devices for arbitrary content in the absence of warrant or suspicion. Come the next terrorist scare, a little push will be all that is needed to curtail or remove the current protections. For example, a scanner that looks only for CSAM images might have its target list extended to scan for other images and its software ``upgraded'' to scan for text as well, at population scale.

\para{Revealing Content Beyond Legitimate Search} CSS systems are designed  to detect material targeted by trusted parties and only reveal this material to law enforcement or intelligence agencies. Both authorized and unauthorized entities may use their vantage points (see \autoref{fig:adversaries}) to re-target the system for other purposes. For instance, the adversary may change the search space to target other content, such as by searching for dissident religious leaders as well as for sex-abuse material. 

An authorized party can expand the legitimate search by demanding that the service provider, or the trusted party composing the list of targeted content or training the machine-learning model, add new content according to their interest. Unauthorized parties could do the same by bribing or coercing staff, or hacking the computers they use (see \autoref{sec:threats-to-css}).

It can be argued that service providers or trusted parties can refuse to include illegitimate target content if it is introduced by adversaries in the updates served to clients. However, if the target material must be kept secret for policy reasons (as is the case with CSAM) then it is unclear how to detect this. Designers---and users---must think carefully about what could stop adversaries from including additional targets in a stealthy manner. 

In the case of CSS based on perceptual hashing, the adversary can launch a second-preimage attack in which they take a legitimate target item (e.g., a CSAM image) and modify it such that the hash of this image also corresponds to other content that they wish to add surreptitiously to the target list (e.g., an LGBTQ+ photo, a photo of a political rival, or a photo of a controversial event such as the Tiananmen protests). We explain these attacks in \autoref{sec:collision}.

In the case of a scanner based on machine learning, the adversary can backdoor the model using specially crafted legitimate target items in such a way that the surreptitiously targeted content also triggers detection.~\autocite{GDG2017}

\para{Revealing Targeted Content to Local Adversaries} The underlying assumption is that a CSS system, like an existing server-based scanner, would only reveal the existence of targeted material to a law enforcement or intelligence agency (as illustrated in \autoref{fig:css_arch}).

However, CSS may also reveal the existence of such material to other recipients who do not have a legitimate need to know. For example, some systems detect nudity. Parents might be informed of LGBTQ+ children sharing nude images before they have come out to their parents, while abusers might be informed that their partners are sharing nude images with other contacts. In such cases, for CSS to reveal even the existence of such material is  a privacy violation. If systems seek to mitigate the risks of intimate partner abuse, they need to be designed carefully for that purpose.

\para{Privacy Harms for Victims}
CSS systems require providers to move parts of their sensitive models or hash lists out of centralized servers, and to deploy them on clients that may be reverse engineered. This increases the probability that adversaries, whether users or others, will  reverse engineer the application to extract the model or list and even extract sensitive information from it. For example, Krawetz has shown that the PhotoDNA perceptual hashes can be decoded to a 26x26 grey-scale thumbnail that permits ``identifying specific people (when the face is the main picture's focus), the number of people in the photo, relative positioning, and other large elements (doors, beds, etc.).''~\autocite{reverse-photoDNA} Apple's current proposal appears much less vulnerable to such original-image extraction, as the NeuralHash output values are shorter.

Research has repeatedly demonstrated that sensitive training inputs can be extracted from machine learning models~\autocite{FredriksonJR15}, and that---even worse---such attacks are quite difficult to prevent.~\autocite{TMWP2020} The only real known defense at present is query control: running the model on a trusted server, observing what queries are made, and stopping the service if a query sequence appears to be trying to extract the model.~\autocite{JSMA2019} Because it would be illegal to provide abuse imagery to attackers and unethical to risk traumatizing abuse victims further, large parts of the models will have to stay in the cloud.

\subsection{Security Risks}

We now consider how a move to client-side scanning may increase security risks. Might an adversary abuse the system to falsely accuse others? And how much more information can an adversary learn about a target by subverting a CSS system running on their device?

\para{Targeting People} A goal of adversaries may be to get particular individuals reported by the system. A state might want to locate dissidents overseas, whether to harass them or perhaps as a first step towards a bigger goal, such as to frame them and then pressure them into becoming an informer for its intelligence service. 

To launch attacks of this kind, the adversary can send victims content that appears innocuous but will trigger reporting (see \autoref{sec:collision}).
People already harass journalists by sending them CSAM and then reporting them to the authorities.
Automated reporting might provide a means to scale up such attacks.
%\end{description}

These attacks can already be carried out with server-side scanning. But moving scanning to the client makes the adversary more powerful in several ways. First, now that algorithms run on the user device, adversaries can study them more closely and experiment to improve their attacks. Second, running the scan on user devices to which service providers regularly send updates gives the adversary more opportunities to tamper with the system. Anyone with access to the targeting pipeline, the provider, the communication, or the device might be subverted by the adversary---whether access is direct (e.g., via infiltration or hacking) or indirect (e.g., by exerting pressure for policies to change).

\para{New Software Security Vulnerabilities} CSS increases the attack surface of users' devices. Currently, devices are compartmentalized, which means that in general apps have no access to other apps' data. This is one of the main ways in which the phones we use now are more secure than laptops. Deploying CSS on devices appears highly likely to move them from this compartmentalized model, and away from accountability in general, and very heavily in the direction of secrecy and intelligence collection. We will discuss this in more detail in section~\ref{sec:css-location}.

Moreover, the CSS mechanisms themselves may have bugs~\autocite{KulshresthaMayer21}; and the server-side systems that update them will have bugs, too. Updates are therefore necessary, but are also a powerful way to scale attacks---an example being the Russian attacks on US agencies through SolarWinds. A criminal adversary who can subvert the update mechanism could use it to install ransomware; the new models of ransomware-as-a-service enable such attacks to be monetized quickly with the help of others. 

\subsubsection{Implementation Decisions' Impact on Security}
\label{sec:css-location}

The security risks posed by CSS will depend on where in the client the scanning is done, who writes the code, and who operates the servers to which the clients report. 

\para{The Location of CSS within the Client} On-device detection can happen in one or more apps, in the operating system, or in middleware. 

If scanning is implemented at the app level, circumvention could be as simple as changing apps (e.g., moving from WhatsApp to Signal). Network effects may prevent the general public from doing this, but motivated groups can and will do so. Such a solution would therefore harm general user privacy without achieving law enforcement's stated goals.

If scanning is implemented in the operating system, the scanner will be in a position to monitor all apps and to control the device itself. It would have the same capabilities as the implants installed by police forces on suspects' phones with warrants, and used by intelligence agencies against targets such as newspaper owners and heads of government. Such a scanner could copy all the data from the device, record passwords, report location, and even turn on the microphone and camera by remote command. With full access to the device, investigators can also do ``cloud forensics'': for example, downloading the cookies used to access email, then downloading all the suspect's email to their own servers for examination. If the scanning code is buggy, it could be exploited remotely to the same effect. 

A warning comes from the first CSS system to be deployed at scale, China's Green Dam censorware, whose installation was mandated on all PCs sold in the country from 2009. This CSS system was advertised as a porn filter, but also searched for target keywords, such as ``Falun Gong,'' and reported them to the authorities. The Green Dam code was buggy, so it not only reported on users but also enabled any website visited by a Green Dam user to take control of the user's PC.~\autocite{WHY2009} 

There is a third option: scanning  implemented as middleware (e.g., Apple's proposal, see~\autoref{sec:apple}). In this case, the scanner does not have unrestricted access to the phone. Thus, exploits like those against Green Dam are much less likely. Users may have some way to escape scanning; for example, they may be able to avoid actions that trigger the middleware.

\para{Code Origin}
Another important question is who writes the CSS code. If scanning is done within apps, it is unreasonable to expect governments or their contractors to be able to write CSS code for each individual application, as their architecture and operation vary tremendously. Given app update cycles and issues of liability, it is unreasonable to expect developers to allow others to tinker with their code.

Trust issues would be even more severe if scanning code provided by governments were to be inserted at the operating-system level. OS vendors have objected very strongly to proposals for government-mandated surveillance functionality, from Microsoft in the 1990s to Apple in 2016. Microsoft in particular has worked hard for two decades to stop third-party code running with system privilege in components such as device drivers (the task is still not complete). Vendors already struggle to patch all the bugs written by their own developers. Before opening their sanctum to defense contractors, OS vendors would need verifiable assurance that the government-provided code contained neither exploitable bugs nor hidden malicious functionality---an assurance that is impossible to provide.\footnote{Nicole Perlroth describes an exchange between James Gosler of Sandia National Labs and Robert Morris, then Chief Scientist of the NSA's National Computer Security Center. Morris was certain that the NSA could find any backdoors in a code base of less than 10,000 lines, but his elite team could not find the traps that Gosler buried in a program of less than 3,000 lines. \parencite[Chapter 7]{Perlroth2020This-is-How-The}} 
%The Chinese Green Dam double use is an examples of the consequences of this inauditability. 

Another route might be for governments to promote a CSS standard. In the twentieth century, government agencies helped develop international standards for wiretapping, but these turned out to be insecure. They created substantial risks and led to several third-party exploits.~\autocite{athens-affair}~\autocite{Cross2010Exploiting-Lawf}~\autocite{rfc3924}~\autocite{Landau2013} A similar attempt to standardize perceptual hashes and machine-learning models is not currently feasible, as there exists no workable technology to standardize. 

\para{The Location of Server Verification}
Another architectural choice is where devices report targeted content. Reporting directly to police stations around the world would introduce uncontrollable security risks. Reporting directly to a single central agency in each country might be more manageable from a technical security viewpoint, but would raise very serious issues of governance and oversight. Those providers who do server-side scanning already have screening infrastructure.~\autocite{FacebookHate}~\autocite{FacebookChild}~\autocite{GoogleTerror} They already report abuse at scale, with teams experienced at assessing disturbing material and contacting the appropriate receiver (local law enforcement, social workers, or even a child's parents). While these mechanisms are far from perfect, and are the subject of justified policy discussion, there is no silver bullet in moving the scanning to the client. The issues around dealing lawfully, effectively, and sensitively with alarms at scale will remain the same.

\section{CSS Is Less Efficacious in Adversarial Environments}\label{sec:effectiveness_attacks}

%\note{There is no threat in the threat model (\autoref{sec:threats-to-css}) for these attacks. Is this a problem? Should we be clearer there, or we just go like this?}

Both distributors and consumers of targeted material may seek to defeat a CSS system by making it useless for enforcement. This can be done in broadly two ways: first, by ensuring that targeted material of interest to them evades detection (i.e., by increasing the rate of false negatives), and second, by tricking the CSS system into flagging innocuous content, thereby flooding it with false alarms (i.e., by increasing the rate of false positives). 

Such attacks are not new. They have been carried out for years on server-side scanners such as spam filters, but a move to client-side scanning brings one telling advantage to adversaries.~\autocite{KulshresthaMayer21} The adversary can use its access to the device to reverse engineer the mechanism. As an example, it took barely two weeks for the community to reverse engineer the version of Apple's NeuralHash algorithm already present in iOS 14, which led to immediate breaches as we explain in \autoref{sec:collision}.

\subsection{Evasion Attacks}\label{sec:evasion}

In evasion attacks, the adversary aims to get targeted content past the scanner. There are simple attacks where consumers disable scanners or avoid using devices that may contain them; here, we consider cases where distributors alter content so as to cause errors in the scanner's target-detection mechanism---both false negatives (where targeted content is passed as innocuous), and false positives (where innocuous content is altered to cause alarms at scale to swamp the defenses). Such attacks have been demonstrated for both perceptual hashing and machine learning. 
%% CT- WHILE NICE, It is not clear that this adds much to the discussion, the ROC is not used anywhere else
%
%The two are in fact linked; the trade-off between false positives and false negatives in a decision problem is known as the \textit{receiver operating characteristic} or ROC. If an attack can degrade the ROC, then there will either be more false positives, or false negatives, or both, depending on how the system responds.\footnote{The ROC is not limited to machine learning; it was developed during World War 2 to analyze the performance of radar in the presence of jamming. It is widely used in electronic warfare and a feature of many other recognition systems, including biometrics.}

\para{Evasion Attacks on Perceptual Hashing} Hao et al.\ showed that it is possible to create images whose perceptual hash is very different from that of another visually similar image.~\autocite{HaoLJW21} They showed this is possible even against server-side scanning, where the adversary does not have access to the scanner, and can submit only limited queries to it. In fact, attacks on perceptual hashing algorithms are highly transferable, in that an image manipulation that confuses one algorithm is likely to confuse many others, too. Hao et al. suggested that a defense would be to add more hashes in the server to make the ensemble robust; and after researchers discovered attacks on NeuralHash, Apple claimed that they intended to follow this strategy.~\autocite{Verge-apple2021} Yet Hao et al. provided examples of algorithms that can be trained to evade combinations of classifiers.

Jain et al. studied evasion in the context of CSS.~\autocite{JainCretudeMonteye} Their results confirmed Hao et al.'s findings: for a large number of images (99.9\% of the images in their study), it is possible to find nearly imperceptible changes to an image that cause it to not be detected any more. For a detector to avoid these false negatives, the number of images flagged would be orders of magnitudes larger, rendering manual review infeasible. Their experiments also showed that effective perturbations span a wide range of modifications, so building a robust defense that blocks all of them appears to be a wicked problem.

\para{Evasion Attacks on Machine Learning} Since 2013, when two independent teams led by Szegedy and Biggio showed how to evade machine-learning classifiers with small perturbations~\autocite{biggio2013evasion,szegedy2013intriguing}, there has been a rapidly growing body of research on the topic.~\autocite{carlini-list} 

In a nutshell, given access to models, or even just to the images or text used to train them, virtually any content can be tweaked to escape detection. In many circumstances, it is possible to create perturbations that evade any model.~\autocite{Moosavi-Dezfooli17} Most of the proposed defenses have either been broken,~\autocite{TramerCBM20} or shown to impose a significant penalty on the model's performance.~\autocite{TsiprasSETM19} Even more damning is increasing evidence that there are fundamental trade-offs that prevent the detection of all kinds of adversarial effects~\autocite{TramerBCPJ20}, and that being able to detect adversarial inputs to filter them may be an unavoidably hard problem.~\autocite{Tramer21}

\para{Evasion Through Poisoning Attacks} Another way to achieve evasion is to \textit{poison} the scanning model (or hash list).
In a poisoning attack, the adversary deliberately influences the training dataset or process to manipulate a model and cause misclassifications.~\autocite{BarrenoNSJT06,BiggioNL12}

Spammers try to poison spam filters by sending spam to email accounts they control, and clicking the ``not spam'' button.  In the context of CSS, poisoning could involve an adversary altering the feed of data to the agency that curates the target list, so that some class of material that should be targeted is passed as innocuous instead; or, as we discuss in \autoref{sec:functionality}, to give a false alarm on innocuous content.

\subsection{False-positive Attacks}\label{sec:collision}
In a false-positive attack, the adversary creates and distributes innocuous content that is falsely detected as targeted, so as to trigger a large number of false alarms. False alarms are a standard way of disabling alarm systems in a wide range of contexts.\autocite{SEv3}

\begin{figure}
\centering
  \includegraphics[width=0.45\textwidth]{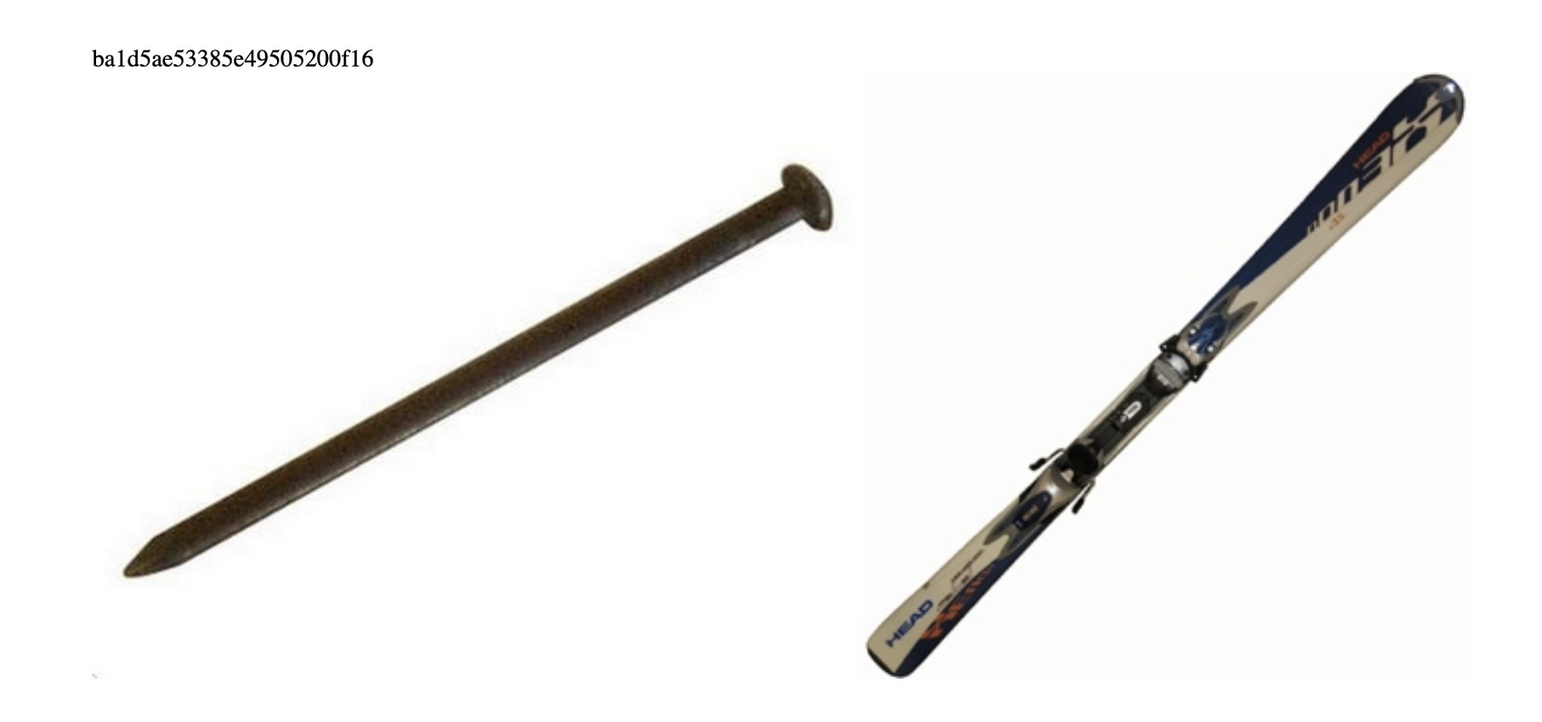}
  \includegraphics[width=0.45\textwidth]{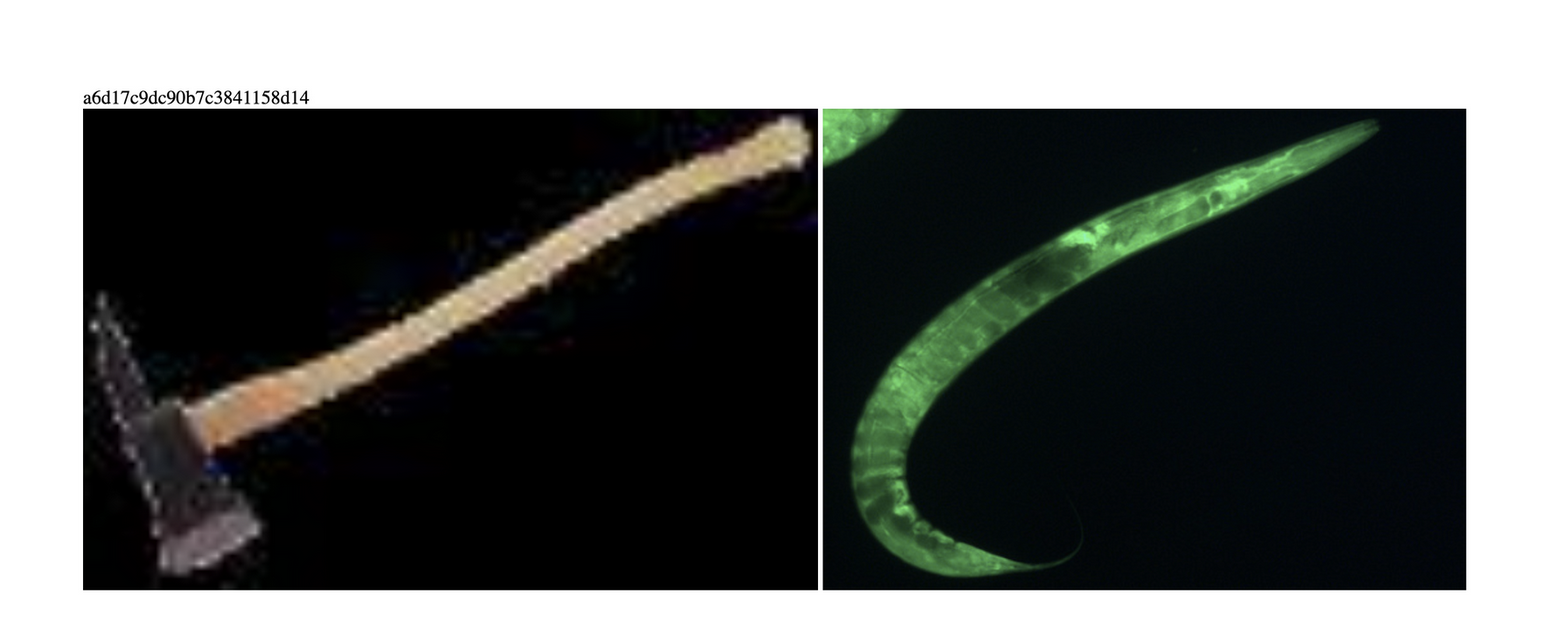}
  \includegraphics[width=0.45\textwidth]{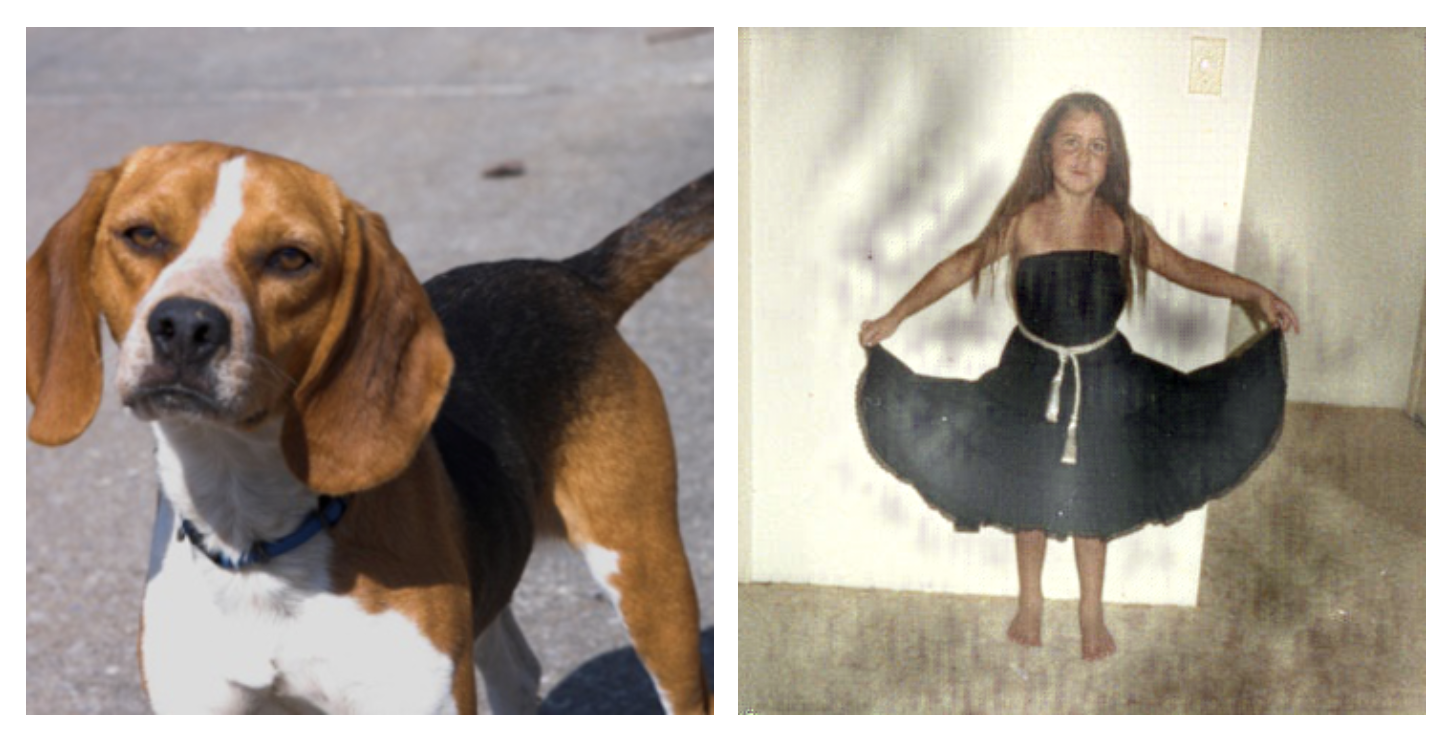}
  \caption{Collisions of the NeuralHash function extracted from iOS 14. 
   \textbf{Top}: Two pairs of accidentally colliding images in the ImageNet database of 14 million sample images; 
  \textbf{Bottom}: An artificially constructed pair of colliding images.}
  \label{fig:collision}
\end{figure}

\para{False-positive Attacks on Perceptual Hashing} When the perceptual hashes of two distinct images match, they are called a collision.\footnote{Some perceptual hash systems, such as Apple's NeuralHash, require that two fingerprints be numerically equal to match, while others, such as Microsoft PhotoDNA, consider two hashes to be a match if they are numerically close.} Perceptual hash functions used in content-scanning systems are designed so that accidental collisions occur with relatively low probability. Yet they still occur in the wild~\autocite{neuralhash-collision}, and we show two examples in the top row of \autoref{fig:collision}. Given that perceptual hash functions are designed so that similar images give similar hashes, such collisions are unavoidable. To create a collision with an existing target image, the adversary has to find what cryptographers call a \textit{second preimage} of the hash (the target image is the first preimage). Generally, finding second preimages is a harder problem than finding collisions, but it still turns out to be easy for many of the known perceptual hash functions. 

The ease of finding second preimages of illicit content opens the door to adversarial collisions, where an attacker creates images with hashes that ``match'' those of targeted images. A user who downloads such content would trigger a match in their CSS system, which could result in a notification to their provider and possibly to law enforcement. Within 48 hours of publication of the NeuralHash code, anonymous parties created multiple second-preimage collisions (see \autoref{fig:collision}, bottom row, for an example) and published a toolkit to generate more.\autocite{neuralhashcoll21} Protesters (or distributors of targeted content) could use this kit to overwhelm Apple's detectors with false positives.

\para{False-positive Attacks on Machine Learning} As we have explained, in a machine-learning environment, it is easy to create two files for which the model outputs the same decision. As we noted, researchers in adversarial machine learning have developed many algorithms to do this, as well as software toolkits that do much of the work.\autocite{papernot2018cleverhans} Many students of computer science and engineering have acquired practical experience of the tools. Although adversarial machine-learning attacks are possible on server-side systems, they are even easier when the scanning results are available on the client, as the adversary can test the model with repeated queries, and perhaps even extract the model. They could either modify targeted images to evade detection, or modify innocuous images to create false alarms.

\para{False-positive Attacks via Poisoning and Backdooring} The techniques in \autoref{sec:evasion} can be used to poison a model or hash list in such a way that benign images trigger the detector. An even more pernicious variant on the same theme is backdooring, where malicious functionality is hidden in a neural network at training time.~\autocite{GDG2017} This can be done in a surprisingly large number of ways and is an active area of current research. The outcome might be racial or gender bias, or to spot a target individual in any photo and then misreport that photo as a specific abuse image. 

\section{Practical Objections to CSS Deployment}\label{sec:practice}

There's an old adage: ``in theory, there is no difference between theory and practice, but in practice, there is.'' This very much applies to CSS. We now look in more detail at the engineering options and the trade-offs between capability, trust and risk.

\subsection{Fairness and Discrimination} 

Content scanners are designed to find approximate matches to targeted files so they can recognize content that has been re-encoded or tweaked. Machine-learning systems are able to identify entirely new content that is somehow similar to known targeted content. In both cases, false positives are inevitable: some innocuous content will be flagged as targeted. The likelihood of such errors increases if there are differences between the distribution of training data and the distribution of data encountered in the wild---a phenomenon known as {\em distributional shift}. This increased error tends to affect particular subsets of the population; and in the case of images, the risk of error typically increases for minorities. In the case of text scanning, we expect the same to hold for minority languages; if the Europol ``bad word list'' contains slang words for drugs and guns in Albanian, but the filters do not have a sizeable corpus of innocuous Albanian text for training, then Albanian speakers might well be wrongly targeted. In both cases, the move from centralized to distributed scanning is likely to make fairness harder to monitor; audit will not only be more difficult technically, but the incentive to do it will be weaker.

Finally, even if false positives occur with low probability, messaging systems operate at hyperscale: many billions of messages are sent each day. If false positives were only one in a thousand, millions of messages would have to be assessed centrally, imposing very high costs on service providers. Existing scanning systems that use machine-learning techniques at this scale end up requiring thousands of human moderators who have to assess many of the worst things that humans can do: not just sex abuse and terrorist recruitment, but animal cruelty, videos of gangland torture and murder, and much else. Their assessments are used not just to train the machine-learning models and decide what gets taken down, but may also be sent to appropriate law enforcement or other agencies. As with child abuse, reality is complex. For example, services that simply take down footage of war crimes end up depriving investigative journalists, human-rights lawyers, and diplomats of the evidence they need to alert the international community and to bring war criminals to justice.

\subsection{Barriers to Scale}

Any engineer who has worked with large-scale systems knows the importance of scaling up testing. First, a prototype is tested on internal users. Then it is tested on successively larger populations to understand error rates, error handling, updates and unexpected dependencies. The only practical way to do things is to evolve systems that work and then write the regulations for their governance. Yet governments often start from the regulations. The history of government IT projects is thus littered with expensive fiascoes, and a particular hazard is for governments to mandate something without knowing whether it will work at scale. That was the rationale behind the 1996 US National Academies study on encryption policy recommendation that aggressive government promotion of escrowed encryption should not be done until after operational experience with escrowed encryption at scale.~\autocite{CRISIS1996} We have a similar lack of experience with the deployment of scanning at scale; there are many unresolved issues with server-side scanning, and a move of scanning to clients would not only create many more dependencies and complexities but also attract determined and skilled adversaries seeking to create various types of failures.

\subsection{Differing Update Cycles}

Software bugs are inevitable, and some of them lead to exploitable vulnerabilities. A significant difference between the computer systems of today compared with those of the last century is the need for regular updates. On complex platforms, such as CSS systems, dozens of vulnerabilities may need to be patched every month. 

The processes by which these vulnerabilities are reported to vendors and patches are shipped is one of the critical factors in practical security engineering. While software vendors mostly offer a ``bug bounty'' or reward to those who report a vulnerability so it can be fixed, much larger rewards are available from cyber-arms manufacturers who use vulnerabilities in hacking tools that they sell to state actors and others.~\autocite{Perlroth2020This-is-How-The}

While Apple fixes most vulnerabilities in iOS and ships updates even for five-year-old iPhones, many Android vendors do not patch phones that are no longer on sale. As a result, most Android phones in use worldwide are insecure; adversaries can take them over using publicly known vulnerabilities~\autocite{TDA2015}. 

CSS systems deployed in Android devices that are not regularly patched must therefore be assumed incapable of providing the secrecy needed to run a CSS system, putting in jeopardy the system's global effectiveness.

\subsection{Jurisdictional Issues}
For general CSS systems, such as those envisaged by the EU to scan multiple formats (images, video streams, and text) looking for varied types of targeted material (CSAM, terrorism, or others), jurisdiction will create thorny issues, just as it did for the attempts in the 1990s to mandate government access to encryption keys. Could a Chinese agent open a free Wi-Fi hotspot in a location where government officials or tech company staff hang out, and route the communications from some of their devices via China using a VPN, so that their CSS systems start operating according to Chinese rules? Operations like this would enable countries to conduct surveillance on other countries and their domestic political rivals, whether at home or in exile. Some authoritarian rulers already falsely denounce rivals as terrorists and put them on the Interpol red list \autocite{Higgins2016How-Moscow-Uses}. CSS for terrorism would provide a ready-made means of repression and political manipulation. 

This problem is made worse by the fact that different governments, including the USA, the EU, and the UK, have different demands. Not only does this variety point to a real lack of agreement on what problem needs solving; it vastly complicates any proposed scanning solution.

Such risks cannot be mitigated through technological means. Service providers deploying CSS can establish policies to try to accommodate conflicting jurisdictional requirements, but in practice will eventually have to comply with the demands of countries in which they have substantial sales or in which they employ staff. 

\subsection{Secrecy Is Incompatible with Accountability} 
Preventing data leakage from CSS systems is fundamentally incompatible with accountability. Existing server-side scanning systems train on abuse data collected by the provider, or obtained under special legal agreements from agencies such as NCMEC. There is little public visibility of this process, which increases the risk of states targeting other material, whether by covert coercion of curators in their own jurisdiction, by hacking curators in other jurisdictions, or by some kind of manipulation or fraud.

CSS is no different. When it is illegal to make models' training data public, as is typically the case with sex-abuse and terrorism material, the risk of training-data extraction means that the models also cannot be public. This makes it extremely difficult to determine what content is being extracted from people's phones. As new targeted content arises all the time and models will need to be updated to target it, surveillance can, by design, evolve and broaden without public oversight. 

If the risk of adversaries extracting targeting information makes it imprudent or even illegal to use CSS on insecure platforms, such as Android phones and Windows devices, what then? Is it proposed to make untraceable wiretapping easy, but only on secure devices such as iPhones? That would motivate people to buy less-secure devices.

\section{CSS Cannot Be Deployed Safely} \label{sec:CSSvsPrinciples}

In this section, we recap our analysis of whether CSS systems are likely to adhere to the security and policy principles discussed in~\autoref{sec:principles}, in the light of the security analysis in~\autoref{sec:functionality} and \autoref{sec:effectiveness_attacks} and the deployment considerations in~\autoref{sec:practice}. Next we analyze Apple's recent CSAM proposal with respect to safety and security.

\subsection{Does CSS Adhere to Security and Policy Principles?}

\para{Core Security Engineering Principles} In this section, we return to Jerry Saltzer and Mike Schroeder's design principles from Section~\ref{sec:secengprin}.

% economy-of-mechanism
From a security-engineering perspective, CSS systems add complexity to already complex systems. Entities other than the platform operator exist within the device's security perimeter. The curator that supplies the target list is trusted, and the same holds for the other parties that supply it in turn. These entities provide critical inputs to the targeting mechanism that they can change at will, and everyone must trust them as a participant within the device's security perimeter. This treads on the economy-of-mechanism principle; it extends the trusted computing base and thus the attack surface while giving no clear benefit to the user.

% fail-safe, economy of mechanism
CSS systems are built with no clear separation of privilege to protect citizens from abuse. The software provider, the infrastructure operator, and the targeting curator must all be trusted. If any of them---or their key employees---misbehave, or are corrupted, hacked or coerced, the security of the system may fail. We can never know when the system is working correctly and when it is not. 

% least-common mechanism
The pervasive deployment of CSS breaks the least-common-mechanism principle. Any security failure in the system could affect every user of the device. 

% least privilege
CSS is at odds with the least-privilege principle. Even if it runs in middleware, its scope depends on multiple parties in the targeting chain, so it cannot be claimed to use least-privilege in terms of the scanning scope. If the CSS system  is a component used by many apps, then this also violates the least-privilege principle in terms of scope. If it runs at the OS level, things are worse still, as it can completely compromise any user's device, accessing all their data, performing live intercept, and even turning the device into a room bug. 

% open design
CSS has difficulty meeting the open-design principle, particularly when the CSS is for CSAM, which has secrecy requirements for the targeted content. As a result, it is not possible to publicly establish what the system actually does, or to be sure that fixes done in response to attacks are comprehensive.\footnote{There are other systems that place these principles under stress. For example, TLS is a complex system with many components that must all be evaluated together. Although these components are the product of competing entities, these entities share a common goal that is aligned with the safety and security of the device's owner. CSS is not. CSS works against the device owner in their otherwise private space.} Even a meaningful audit must trust that the targeted content is what it purports to be, and so cannot completely test the system and all its failure modes. 

% psychological acceptability
Finally, CSS breaks the psychological-acceptability principle by introducing a spy in the owner's private digital space. A tool that they thought was theirs alone, an intimate device to guard and curate their private life, is suddenly doing surveillance on behalf of the police. At the very least, this takes the chilling effect of surveillance and brings it directly to the owner's fingertips and very thoughts.

We conclude that the requirements and constraints of CSS systems are at odds with important security engineering practices. CSS systems are by construction untrustworthy, and vulnerable in adversarial environments. 

\para{Core Policy Principles} Here, we return to the Carnegie principles discussed in Section~\ref{sec:carnegie}. It is unclear that CSS systems can fulfill the Law Enforcement, Utility, and Equity principles. We have shown several ways in which adversaries can jam and evade scanning, regardless of the underlying technologies; and that both natural and adversarial errors may exacerbate inequity. Any CSS proposal would require careful study and testing to determine the ease with which false positives and false negatives could be found, and how easily malicious actors could use them to deny service, or to frame or blackmail innocent parties.

When CSS is deployed on all devices rather than only on those of suspects, it breaks the Authorization principle. Specificity fails because the need to update target lists to keep up with new abuses enables those updates to expand the nature and scope of targeting and to allow the initial mission to expand into new ones. If a government decides to identify dissidents, for example, all that may be involved is adding images of protest figures such as religious leaders to the target list.

Finally, most arguments for CSS are justified by narratives about target material that cannot for legal reasons be made public (such as images of sex abuse and politicized murders). The resulting secrecy makes the Auditability principle hard to fulfill. Secrecy also makes it harder for CSS to abide properly by the Transparency, Evaluation, and Oversight principles.

We conclude that the architecture of CSS systems makes them insecure and potentially ineffective in undetectable ways. It also makes it difficult for designers and service providers who might deploy them to abide by the Carnegie principles themselves. It is therefore impossible to ensure that CSS will be deployed judiciously, and it cannot reasonably be claimed that the risk such a deployment poses to society is necessary and proportionate.

\subsection{Example: An Analysis of Apple's August 2021 Proposal}
\label{sec:apple}

In August 2021, Apple proposed the first production CSS system, which had the potential to be deployed at global scale and installed in more than a billion Apple devices. Apple's proposal was primarily intended to detect CSAM using an on-device detection component that works with Apple servers using cryptographic protocols. Apple designed the scanning components to operate on photos stored in the Camera Roll (the device photo library) only when the ``iCloud Photos'' cloud synchronization service is turned on.\footnote{Apple made a total of three simultaneous announcements that may have confused the public debate. The second was a machine-learning-based nudity detector designed to notify users under age eighteen if they seemed to be about to send or receive a naked picture (and, if they were under thirteen, to inform their parents). The third was changes to Siri designed to prevent voice-based searches for CSAM material.}

In a nutshell, Apple's proposal follows the CSS operation flow described in \autoref{sec:flows} with two main differences. 

First, in Step 3, targeted content is only added to the hash list sent to the users' devices if two curators in different jurisdictions approve the content.

Second, in Step 4, the scanning step on the device is augmented with advanced cryptography to ensure that matches can only be detected by the server (rather than the client) and only if multiple\footnote{This threshold has currently been set to thirty.} different matches are detected on the device.~\autocite{Federighi2021} When content is detected by the server, the system automatically reveals a low-resolution version of the detected image. As a further protection, the server runs a second perceptual hash function on this material and allows for human review, in order to reduce the impact of false positive matches once the threshold is met.

Apple describes the design goals for their system in a threat model document~\autocite{Apple2021ThreatModel}. First, they seek to create a separation of privilege, so that users should not have to trust Apple or any one sovereign state. Apple will have only one global list of target images, and an image hash will appear on the list only if it is supplied by abuse organizations in two separate jurisdictions, such as NCMEC in the USA and the IWF in the UK. Second, Apple proposes to require multiple matches, followed by human review and additional protections, to avoid accidental false positives. Finally, they propose {\em auditability} as a design goal: this ensures that Apple cannot change the scan database or insert unknown content into it. The latter properties seem poorly specified, but Apple suggests that they may be enforced with the assistance of some trusted third party organization (as yet unspecified by Apple) that can verify the contents of Apple's scanning database. In the next section, we discuss each of these proposed design goals.

\medskip \noindent
{\em Separation of privilege for database content.} While Apple's decision to include only files attested to by multiple child safety organizations raises the bar for attack, this approach depends on Apple being able and willing to enforce this policy. It could stop enforcing this policy locally or globally, whether by a company decision or under pressure from states wishing to maintain sovereignty within their borders. Apple has yielded to such pressures in the past, such as by moving the iCloud data of its Chinese users to three data centers under the control of a Chinese state-owned company~\autocite{Nicas2021}, and by removing the ``Navalny'' voting app from its Russian app store.~\autocite{NavalnyApp}

Moreover, requiring two jurisdictions does not prevent attacks. For instance, in Apple's new system a third party might be able to verify that the targeted images are the intersection of the databases of two other parties. But even if Apple can prevail over the sovereignty of nation states, nations can and do collude. For example, what would happen if Russia and Belarus were to submit a list of hashes? Should Apple accept the list, or reject it and redefine the meaning of jurisdiction? 

\medskip \noindent
{\em Human review to detect false positives.} 
Apple's system uses a new and proprietary perceptual hash function it calls NeuralHash. NeuralHash is designed to concisely ``fingerprint'' images, producing an extremely succinct digest for each scanned image: the remainder of the scanning system is designed to compare hashes on a device to those calculated based on a target database. As we noted, perceptual hash functions may be vulnerable to adversarially generated preimages that cause content to be inaccurately classified. A pre-release version of NeuralHash was duly recovered in August 2021, and researchers have now published tools to construct images that produce the same fingerprint, and thus will appear identical to Apple's scanning system\autocite{neuralhash-collision}.

Apple's design aims to reduce the impact of such false positives, using three countermeasures: $(1)$ Apple will not be alerted unless at least thirty positives occur on a given device; $(2)$ colliding images will also be compared, using a second perceptual hash function that will be kept secret and run on Apple's servers; and $(3)$ all reported positives will be screened by a human reviewer employed by Apple, rather than triggering automated law-enforcement reporting. Just as pressure from governments could push Apple to target other photos than sex abuse, so the threshold of thirty targeted photos for human review could be reduced. In short, it would be a small change to transform the scanner on your device to report any targeted images, regardless of any intent to share them or even back them up to a cloud service.

In summary, Apple has devoted a major engineering effort and employed top technical talent in an attempt to build a safe and secure CSS system, but it has still not produced a secure and trustworthy design.

\section{Conclusions and Recommendations}

CSS has been promoted as a magical technological fix for the conflict between the privacy of people's data and communications and the desire by intelligence and law enforcement agencies for more comprehensive investigative tools. A thorough analysis shows that the promise of CSS solutions is an illusion.

Technically, moving content scanning from the cloud to the client empowers a range of adversaries. It is likely to reduce the efficacy of scanning, while increasing the likelihood of a variety of attacks. 

Economics cannot be ignored. One way that democratic societies protect their citizens against the ever-present danger of government intrusion is by making search expensive. In the US, there are several mechanisms that do this, including the onerous process of applying for a wiretap warrant (which for criminal cases must be essentially a ``last resort'' investigative tool) and imposition of requirements such as ``minimization'' (law enforcement not listening or taping if the communication does not pertain to criminal activity). These raise the cost of wiretapping.\footnote{The average cost of a wiretap in 2020 was \$119,000.~\autocite[Table 5]{WiretapReport2020}}

By contrast, a general CSS system makes all material cheaply accessible to government agents. It eliminates the requirement of physical access to the devices. It can be configured to scan any file on every device. And it has become part of some agencies' vision. GCHQ's pitch document ``AI for national security: online safety''~\autocite{gchq2021} sets a goal of:

\begin{quotation}Providing tools and techniques to identify potential grooming behavior within the text of messages and in chat rooms; highlighting the exchange of illegal images and tracking the disguised identities of offenders across multiple accounts; searching out and discovering hidden people and illegal services on the dark web. AI could also enable us to help law enforcement infiltrate rings of offenders and bring them to justice.
\end{quotation}

So the filter code in your phone won't just be looking for illegal pictures. GCHQ goes on:

\begin{quotation}AI tools can also be trained to analyse seized and intercepted imagery, messages, other forms of internet content, and chains of contact, to support investigators in the identification of victims and discovery of accomplice offenders. AI running across both content and metadata could also protect our analysts from unnecessary exposure to traumatically disturbing material.
\end{quotation}

It is unclear whether CSS systems can be deployed in a secure manner such that invasions of privacy can be considered proportional. More importantly, it is unlikely that any technical measure can resolve this dilemma while also working at scale. If any vendor claims that they have a workable product, it must be subjected to rigorous public review and testing before a government even considers mandating its use.

This brings us to the decision point. 
The proposal to preemptively scan all user devices for targeted content is far more insidious than earlier proposals for key escrow and exceptional access. Instead of having targeted capabilities such as to wiretap communications with a warrant and to perform forensics on seized devices, the agencies' direction of travel is the bulk scanning of everyone's private data, all the time, without warrant or suspicion. That crosses a red line. Is it prudent to deploy extremely powerful surveillance technology that could easily be extended to undermine basic freedoms?

Were CSS to be widely deployed, the only protection would lie in the law. That is a very dangerous place to be. We must bear in mind the 2006 EU Directive on Data Retention, later struck down by the European Court of Justice, and the interpretations of the USA PATRIOT Act that permitted bulk collection of domestic call detail records. In a world where our personal information lies in bits carried on powerful communication and storage devices in our pockets, both technology and laws must be designed to protect our privacy and security, not intrude upon it. Robust protection requires technology and law to complement each other. Client-side scanning would gravely undermine this, making us all less safe and less secure.

\medskip
{\bf Acknowledgements:} We wish to acknowledge the contributions of John Gilmore, Matt Green, Mike Specter and Danny Weitzner, who participated in early discussions of this text. We also wish to thank Nicolas Papernot, Ilia Shumailov, Nicholas Boucher and Sam Ainsworth who gave us valuable feedback on early drafts of the sections on machine learning and system security. Finally we would like to thank Beth Friedman, whose copyediting contributed to the clarity of the ideas as well as the clarity of expression. 

\printbibliography

\newpage
\section*{Authors}
\newcommand{\authbio}[1]{\par\medskip\noindent\textbf{#1}}

\authbio{Hal Abelson}
is the Class of 1922 Professor of Computer Science and Engineering in the Department of Electrical Engineering and Computer Science at Massachusetts Institute of Technology.

\authbio{Ross Anderson}
is Professor of Security Engineering at the University of Cambridge and at the University of Edinburgh.

\authbio{Steven M. Bellovin}
is the Percy K. and Vida L.W. Hudson Professor of Computer Science
at Columbia University and affiliate faculty at Columbia Law School.

\authbio{Josh Benaloh}
is the Senior Principal Cryptographer at Microsoft Research and an Affiliate Professor in the Paul G. Allen School of Computer Science and Engineering at the University of Washington.

\authbio{Matt Blaze}
is a Professor of Law; Robert L. McDevitt, K.S.G., K.C.H.S. and Catherine H. McDevitt L.C.H.S. Chair, Department of Computer Science
at Georgetown University.

\authbio{Jon Callas}
is the Director of Technology Projects at the Electronic Frontier Foundation.

\authbio{Whitfield Diffie}
is the Chief Security Officer of Sun Microsystems, Retired.

%\authbio{Matthew Green}
%is an Associate Professor at the Johns Hopkins Information Security Institute.

\authbio{Susan Landau}
is Bridge Professor of Cyber Security and Policy at The Fletcher School and at the School of Engineering, Department of Computer Science, at Tufts University.

\authbio{Peter G. Neumann}
is Principal Scientist in the Computer Science Laboratory at SRI International.

\authbio{Ronald L. Rivest}
is an Institute Professor at Massachusetts Institute of Technology.

\authbio{Jeffrey I. Schiller}
is Enterprise Architect at Massachusetts Institute of Technology.

\authbio{Bruce Schneier}
is a fellow and Lecturer at Harvard Kennedy School, a fellow at the Berkman Klein Center for Internet \& Society at Harvard University, and Chief of Security Architecture at Inrupt, Inc.

\authbio{Vanessa Teague}
is CEO of Thinking Cybersecurity and an Associate Professor (Adj.) at the Research School of Computer Science at the Australian National University.

\authbio{Carmela Troncoso}
is an Assistant Professor at \'Ecole Polytechnique F\'ed\'erale de Lausanne (EPFL).

\end{document}